\documentclass[aps,prd,twocolumn,showpacs,preprintnumbers,amsmath,amssymb,nofootinbib,scrartcl]{revtex4-1}

\usepackage{color}
\usepackage{graphics} 
\usepackage{graphicx}
\usepackage{epsfig} 

\begin{document}

\title{General treatment for dark energy thermodynamics}

\author{H. H. B. Silva$^{1}$}
\email{heydson@dfte.ufrn.br}

\author{R. Silva$^{1,2}$}
\email{raimundosilva@dfte.ufrn.br}

\author{R. S. Gon\c{c}alves$^{3}$}
\email{rsousa@on.br}

\author{Zong-Hong Zhu$^{4}$}
\email{zhuzh@bnu.edu.cn}

\author{J. S. Alcaniz$^{3}$}
\email{alcaniz@on.br}

\affiliation{$^{1}$Universidade Federal do Rio Grande do Norte, Departamento de F\'{\i}sica, Natal - RN, 59072-970, Brasil}

\affiliation{$^2$Universidade do Estado do Rio Grande do Norte, Departamento de F\'{\i}sica,  Mossor\'{o}--RN 59610-210, Brasil}

\affiliation{$^{3}$Observat\'orio Nacional, 20921-400 Rio de Janeiro - RJ, Brasil}

\affiliation{$^4$Department of Astronomy, Beijing Normal University, Beijing 100875, China}


\date{\today}

\begin{abstract}
In this work we discuss a general approach for the dark energy thermodynamics considering a varying  equation of state (EoS) parameter of the type $\omega(a)=\omega_0+F(a)$ and taking into account the role of a non-zero chemical potential $\mu$. We derive generalized expressions for the entropy density, chemical potential and dark energy temperature $T$ and use the positiveness of the entropy to impose thermodynamic bounds on the EoS parameter $\omega(a)$. In particular, we find that a phantom-like behavior $\omega(a)< -1$ is allowed only when the chemical potential is a negative quantity ($\mu<0$).
\end{abstract}

\maketitle

\section{Introduction}

The discovery of the cosmic acceleration is certainly one of the most important cosmological results of the last decades. It implies that either the Universe is currently dominated by an exotic dark energy component or that the Einstein's general theory of relativity breaks down at very large scales \cite{review}. The profound implications of this discovery have motivated observational efforts to understand the mechanism behind the cosmic acceleration, with a
number of experiments that aim to map the history of expansion and growth of structure with percent-level precision being planned for the next years. From the physical point of view, several fundamental questions on the origin and nature of the dark energy still remain completely opened. Certainly among these, the thermodynamical behavior of this component plays an important role not only in constraining its time evolution \cite{PotNulo} but also in
predicting the dynamical and thermodynamical fate of the Universe \cite{alcaniz2004}.

The thermodynamic behavior of the dark energy considering a constant equation-of-state parameter (EoS) $\omega$ with null chemical potential ($\mu=0$) was discussed in \cite{alcaniz2004}. The authors found that not only a phantom behavior ($\omega < -1$) is forbidden by the
second law of thermodynamics but also that the temperature associated to this component increases with the cosmic expansion (see also \cite{gonzalez}). Later on, several authors~\cite{ademir2008,ademir2008-2} extended these results considering a non-zero chemical potential $\mu$. They showed that for values of $\mu < 0$, the EoS parameter may cross the so-called phantom divide line from which the fluid acquires a phantom-like behavior. More recently, we studied some thermodynamic aspects of a dark energy component described by a varying EoS parameter~\cite{PotNulo}. We found that the time-dependent term behaves like a bulk viscosity in the temperature evolution law and placed some thermodynamic bounds on the parameters $\omega_0$ and $\omega_a$ from the second law of thermodynamics and from the positiveness of the entropy -- where we have assumed the statistical microscopic concept of entropy, i.e. $S=k_B\ln W>0$.

In this paper, we develop a general treatment for the dark energy thermodynamics  considering (i) non-zero chemical potential and (ii) a time-dependent EoS parameter. {Without loss of generality}, we parameterize the function $\omega(a)$ as
\begin{equation} \label{pb}
{\omega(a) \equiv {p_x \over \rho_x} =  w_0+F(a)}
\end{equation}
with encompasses most of the relevant cases discussed in the literature, e.g., ${F(a) = \omega_a(1 - a)}$\cite{cpl1,cpl2}; ${F(a) = - \omega_a \ln{a}}$ \cite{log}; ${F(a) = \omega_a(1-a)/(1-2a+2a^2)}$ \cite{barboza08} (see also \cite{johri} for other parameterizations).

From this background, new evolution laws of some physical quantities
such as entropy density and chemical potential are obtained and new
constraints are imposed to $\omega(a)$ through the positiveness of
the entropy. Therefore, considering this
thermodynamic condition, we also find that, although values $\mu <
0$ are compatible with phantom models without the need to consider
negative temperatures (as argued by \cite{ademir2008,ademir2008-2}),
for $\mu \geq 0$ phantom cosmology becomes physically meaningless.
We also show that most of the results discussed earlier in the
literature are particular cases of the treatment discussed here.
Throughout this paper a subscript $0$ stands for
present-day quantities and a dot denotes time derivative. We use the
speed of light as $c=1$.

\section{Thermodynamics of dark energy}

Let us consider a flat, homogeneous and isotropic cosmological model
described by the Friedmann-Robertson-Walker (FRW) metric:
\begin{equation}
ds^2=dt^2-a^2(t)(dr^2+r^2d \theta^2 +r^2sin^2 \theta d \phi^2),
\end{equation}
In this scenario, the dark energy component is considered as a
relativistic perfect fluid, in which its equilibrium thermodynamic
states are described by an energy momentum tensor
\begin{equation} \label{tens}
T^{\delta \nu}=\rho_x u^{\delta} u^{\nu} - p_x h^{\delta \nu}\;,
\end{equation}
a particle current,
\begin{equation} \label{c-part}
N^{\delta}=nu^{\delta}\;,
\end{equation}
and an entropy current,
\begin{equation} \label{c-ent}
S^{\delta}=s u^{\delta}=n\sigma u^{\delta}\;,
\end{equation}
where $h^{\delta \nu}\equiv g^{\delta \nu}-u^{\delta} u^{\nu}$ is
the usual projection tensor; $n$ is the particle number density; $s$
and $\sigma$ are the entropy density and the specific entropy (per
particle) respectively \cite{landau}. The conservation of $T^{\delta
\nu}$ and $N^{\delta}$ provides the following relations
\begin{equation}\label{enmon}
u_{\nu} T^{\delta \nu};_{\delta}= {\dot \rho_x} + 3 (\rho_x +
p_x)\frac{\dot a}{a} = 0\;,
\end{equation}
\begin{equation} \label{nalpha}
N^{\delta};_{\delta}= \dot n + 3 n \frac{\dot a}{a}  = 0\;,
\end{equation}
where semi-colons mean covariant derivative. Integrating the above
expressions, it is straightforward to show for a general $\omega(a)$
that
\begin{equation} \label{rho}
\rho_{x } = \rho_0 \left(\frac{a}{a_0}\right)^{-3}
\exp\left[{-3\int{\frac{\omega(a)}{a}}da}\right]\;,
\end{equation}
\begin{equation} \label{n}
n = n_0 \left(\frac{a}{a_0}\right)^{-3} \;.
\end{equation}

In order to derive the evolution law of dark energy temperature,
first we consider that the physical quantities $p_x$, $\rho_x$, $n$
and $\sigma$ are related to the temperature $T$ trough the Gibbs
law: $nTd\sigma = d\rho_x - {{\rho_x + p_x} \over n}dn.$ Assuming that $\rho_x = \rho_x(n,T)$ and $p_x = p_x(n,T)$, it can be shown
that such evolution is given by \cite{weinberg,silva,silva2}
\begin{equation} \label{evol-t}
{\dot T \over T} = \biggl({\partial p_0 \over \partial
\rho_x}\biggr)_{n} {\dot n \over n} + \biggl({\partial \Pi \over
\partial \rho_x}\biggr)_{n} {\dot n \over n} \;,
\end{equation}
where the dark energy pressure is composed of a constant term
($p_0$) and by another time-dependent term ($\Pi$) as follow
\begin{equation} \label{px}
p_x \equiv  p_0 + \Pi = \omega_0\rho_x +
F(a)\rho_x=\omega(a) \rho_x\;.
\end{equation}
The above equations provide the relation
\begin{equation} \label{temp}
T = T_0 \; \exp\left[{-3\int{\frac{ \omega(a)}{a}}da}\right]\;.
\end{equation}
It should be noted that the dark energy temperature is always
positive and growing in the course of the universe expansion
regardless of the $\omega(a)$ value (because the EoS parameter is a
negative quantity). A likely physical meaning for this behavior is
that thermodynamic work is being done on the system through an
adiabatic expansion \cite{alcaniz2004} (For a alternative
thermodynamics approach which the fluid takes temperature negative
values see, e.g., \cite{gonzalez,thermo2}).

On the other hand, Eq. (\ref{enmon}) can be rewritten as
\begin{equation}
{\dot\rho_x} + 3(\rho_x + p_0) \frac{\dot a} {a}  = -3 \Pi \frac{\dot a}
{a}.
\end{equation}
This means that the varying part of the dark energy pressure
$\Pi$ plays the role of an entropy source term and, therefore, the
$\omega(a)$-fluid description assumed here mimics a fluid with bulk
viscosity. It suggests a more complete form of entropy density
evolution which is different from the widely used expression
\cite{alcaniz2004,ademir2008,ademir2008-2}. Our starting point is
the well-known Euler relation: $Ts=\rho_{x} + p_{x} - \mu n$. From
equations (\ref{rho}), (\ref{temp}) and the comoving volume $V
\propto {a}^3$, we find
\begin{equation} \label{sigma}
\sigma \equiv \frac{S}{N}= \frac{\rho_0} {T_0}\frac{V_0} {N}
\;[1+\omega(a)]-\frac{\mu_0}{T_0}\;,
\end{equation}
where we used the condition $\mu/T=\mu_0/T_0$ which will be
demonstrated below using the definition of Gibbs free energy.
Therefore, using the above relation and taking the covariant
derivative of Eq. (\ref{c-ent}) we obtain
\begin{equation} \label{c-ent2}
S^{\alpha};_{\alpha}=\dot s + 3s \frac{\dot a}{a} = \frac{\rho_0}
{T_0}\frac{V_0} {V} \; \dot \omega(a) \;,
\end{equation}
or still
\begin{equation} \label{diff}
\frac{ds}{da}+\frac{3s}{a}=\frac{\rho_0} {T_0} \frac{V_0}{V}\omega'(a).
\end{equation}
whose solution is
\begin{equation} \label{entrop}
s =
a^{-3}\left(s_0+\frac{\rho_0}{T_0}[\omega(a)-\omega_0]\right).
\end{equation}
Note that for a constant EoS parameter (${\omega(a)\rightarrow
\omega_0}$) we recovered the entropy relation derived in Refs.~\cite{alcaniz2004,ademir2008,ademir2008-2}.

\begin{figure*}[t]
\centerline{
\includegraphics[width=17cm,height=9.5cm,angle=0]{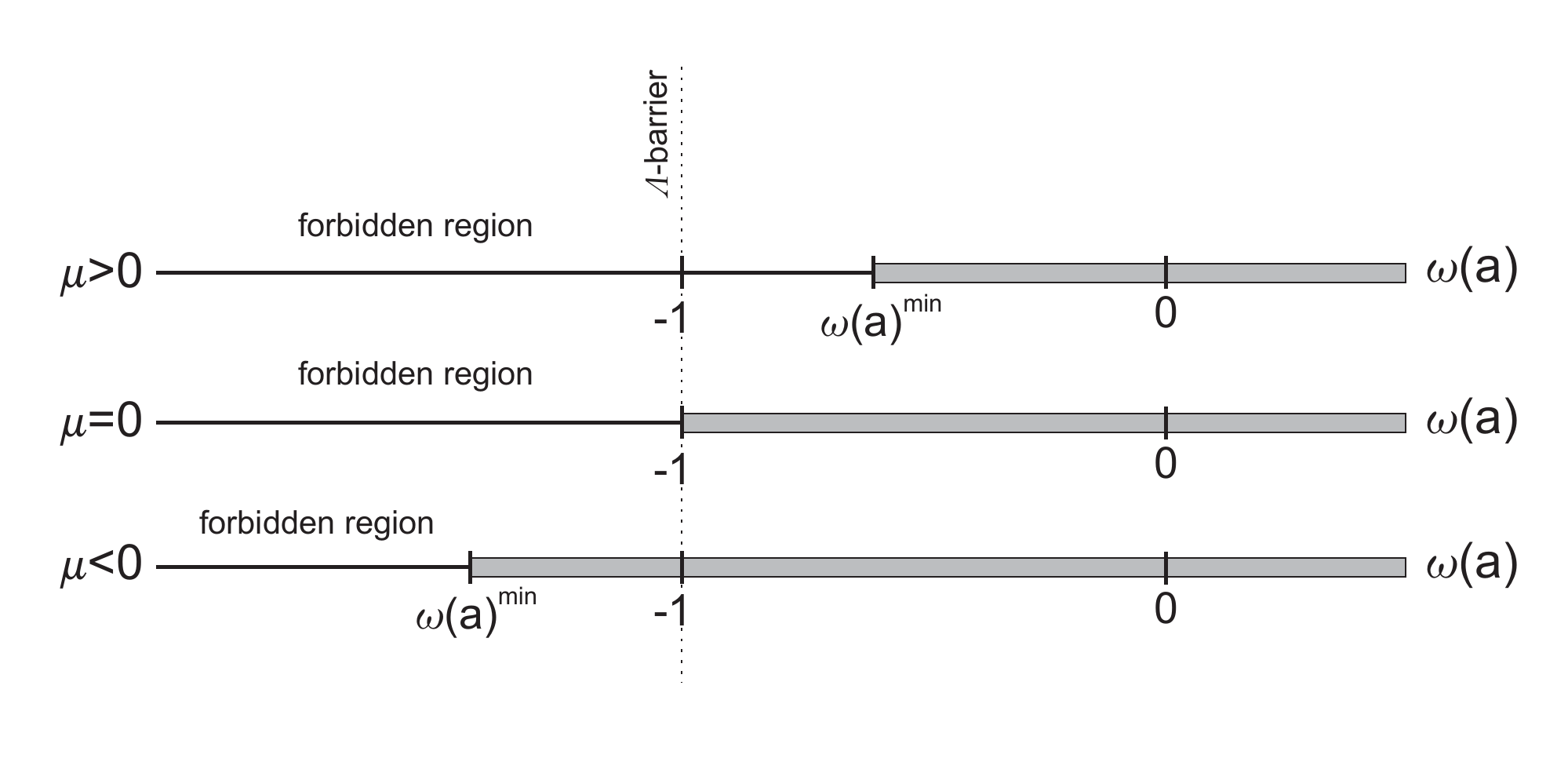}}
\caption{Representation of thermodynamic constraint [Eq.
(\ref{omegaa})] for the cases $\mu<0$,
$\mu=0$ and $\mu>0$. The light gray bars represent the values
allowed for the parameters in order to satisfy the positiveness of
the entropy. The black lines represent the values that do not obey this latter
condition (forbidden region) whereas the dashed line represents the cosmological constant limit. Note that for $\mu<0$ the
phantom-like regime is allowed, but for $\mu=0$ and $\mu>0$, it is
forbidden.}
\end{figure*}

Another relevant physical quantity in thermodynamics analysis is the
chemical potential $\mu$. A way to include it in our treatment is
using the Gibbs free energy \cite{callen}
\begin{equation} \label{gibbs}
G(T,p,N)\equiv U+pV-TS
\end{equation}
with
\begin{equation}
\mu=\left(\frac{\partial G}{\partial N}\right)_{T,p}\;.
\end{equation}
Thus, using the equations (\ref{rho}), (\ref{n}), (\ref{temp}) and
(\ref{entrop}) we find that
\begin{equation}
G=N \left[\frac{\rho_0}{n_0}(1+\omega_0)-\frac{T_0
s_0}{n_0}\right]\frac{T}{T_0}\;,
\end{equation}
and consequently
\begin{equation} \label{mu}
\mu=\mu_0 \left(\frac{T}{T_0}\right)= \mu_0 \exp\left[{-3\int{\frac{
\omega(a)}{a}}da}\right]\;,
\end{equation}
where
\begin{equation}
\mu_0\equiv \left[\frac{\rho_0}{n_0}(1+\omega_0)-\frac{T_0
s_0}{n_0}\right]
\end{equation}
is the present-day value of the chemical potential. Note that the
temperature and chemical potential evolve in the same way (less than
a constant) regardless if the dark energy EoS is constant or
time-dependent. This is so because for both treatment the condition
$\mu/T=\mu_0/T_0$ is always kept
\cite{ademir2008,ademir2008-2,thermo3}.

\section{Thermodynamic Constraints}

The thermodynamics formalism developed in the previous section leads
us to ask the following questions: which constraints can be imposed
to the EoS parameter and the chemical potential in order to satisfy
the positiveness of the entropy? Is the phantom-like behavior of the
dark energy fluid allowed? If so, for which parameter interval?

Taking the above relations we write the total entropy as
\begin{equation}
S=\left(\frac{\rho_0[1+\omega(a)]-\mu_0 n_0}{T_0}\right) V_0 \;,
\end{equation}
whose positivity requires that
\begin{equation} \label{omegaa}
{\omega(a)\geq \omega(a)^{min}=- 1 + \frac{\mu_0 n_0}{\rho_0}.}
\end{equation}
Note that in the limit $\omega(a)^{min} \rightarrow \omega^{min}_0$ we recover the results for constant EoS \cite{ademir2008,ademir2008-2}. 	
Furthermore, we have a limit for the time-dependent EoS parameter,
below which the entropy becomes negative (see, e.g,
\cite{thermo,thermo1,thermo2,thermo3} for other thermodynamic
analyses).

The minimal value $\omega(a)^{min}$ depends explicitly on
the chemical potential. Therefore, firstly we consider
$\mu=\mu_0=0$. In this case, we have ${\omega(a)^{min}=-1}$
and a phantom-like behavior of the dark energy fluid is forbidden.
On the other hand, for $\mu > 0$, the minimal value does not reach
the cosmological constant point ${\omega(a)=\omega_0 =-1}$
and again a phantom-like regime is forbidden. Finally, if the
chemical potential is negative, values of ${\omega(a) < -1}$
are allowed. {A graphical representation of these bounds is
shown in Fig. 1. In particular, in the limit
$\omega(a)\rightarrow \omega_0$, the results
obtained in Ref. \cite{ademir2008,ademir2008-2} are readily recovered}.

\begin{table}[t]  \label{tabela1}
\caption{Thermodynamic
constraints on $\omega(a)$. All particular cases are shown for the different values of the
temperature ($T$), the chemical potential ($\mu$) and of the function  $F(a)$.}
\begin{ruledtabular}
\begin{tabular}{lcccc}
Reference & $T$ & $\mu$ & $F(a)$ & $\omega(a)^{min}$  \\
\hline \\
This work & $>0$ & $\neq 0$ & $\neq 0$ &$-1 +\frac{\mu_0
n_0}{\rho_0}$
\\ \\
\cite{PotNulo} & $>0$ & $=0$ & $\neq 0$ & $-1$
\\ \\
\cite{thermo3} & $<0$ & $\neq 0$ & $\neq 0$ & $-$
\\ \\
\cite{thermo2} & $<0$ & $=0$ & $=0$ & $-$
\\ \\
\cite{ademir2008,ademir2008-2} & $>0$ & $\neq 0$ & $=0$ &$-1 +
\frac{\mu_0 n_0}{\rho_0}$
\\ \\
\cite{gonzalez} & $<0$ & $=0$ & $=0$ & $-$
\\ \\
\cite{alcaniz2004} & $>0$ & $=0$ & $=0$ & $-1$ \\
\end{tabular}
\end{ruledtabular}
\end{table}

In Table I we present  the constraints obtained
through positivity of entropy. Note that most of the
results presented in the literature are particular cases of Eqs.
(\ref{omegaa}). For instance, for $\mu=\mu_0=0$ we recovered the
results of \cite{PotNulo} whereas if one assumes
${\omega(a)\rightarrow \omega_0}$ we obtain the results of
\cite{ademir2008,ademir2008-2}; Finally, considering $\mu=\mu_0=0$
and ${\omega(a)\rightarrow \omega_0}$ we recovered the
results of \cite{alcaniz2004}. Note also that the results obtained
by \cite{gonzalez,thermo2,thermo3} assuming $T < 0$, can also be
mapped in the present investigation.

\section{Conclusions}

We have discussed a general treatment for the dark energy
thermodynamics by considering a varying EoS parameter and non-zero
chemical potential. We have derived general equations for the dark
energy entropy and chemical potential and imposed thermodynamic
constraints for the $\omega$-parameter in order to satisfy the
positiveness of the entropy. These bounds are displayed in Figure 1,
where it is clear that, as the minimal value
${\omega(a)^{min}}$ depends explicitly upon the chemical
potential, only the case $\mu < 0$ allows the phantom-like behavior
without considering negative temperature (as discussed by
\cite{gonzalez,thermo3}). The treatment discussed here
generalizes most of the results investigated in the literature, as
summarized in Table I. Finally, we emphasize that this procedure can
be extended to include non-minimal coupling between dark matter and
dark energy. This issue will be addressed in a forthcoming
communication.


\begin{acknowledgments}

H.H.B.S., R.S., R.S.G. and J.S.A. are very grateful to FAPERJ, CNPq and CAPES for the grants under which this work was carried out. Z.-H.Z. is supported by Chinese Ministry of Science and Technology National Basic Science Program (Project 973) under Grant No. 2012CB821804 and National Natural Science Foundation of China under Grants Nos. 11073005 and 11373014.
\end{acknowledgments}

\end{document}